# Proximity Effect in Periodic Arrays of Superconducting Nanoislands on Thin Graphite Layer

**Yuri Latyshev[1, 2], Anatoly Smolovich [1, *], Andrey Orlov[1, 2], Aleksei Frolov[1, 3], Vyacheslav Vlasenko[4]**

[1]Kotel'nikov Institute of Radio Engineering and Electronics of the RAS, Russia
[2]Laboratoire National des Champs Magnétiques Intenses, France
[3]Moscow Institute of Physics and Technology, State University, Russia
[4]OPTEC LLC, Russia
*Corresponding Author: papa@petersmol.ru

**Abstract**  The regular structure of superconducting nanoislands of alloy W-Ga-C was fabricated on nanothin graphite using focused ion beam. The resistance vs temperature dependence down to 1.7K and the magnetoresistance in field up to 24T were measured both for the bridge containing nanoislands and for the reference bridge without islands. The difference between those measurements demonstrates the proximity effect on a regular structure of superconducting W-Ga-C nanoislands on nanothin graphite layer.

## 1. Introduction

Graphene is a material which has an unique electronic spectrum. Carriers in graphene are massless chiral Dirac fermions. Since the discovery of graphene in 2004 [1] electronic structure of graphene was studied in detail [2-4]. Today hybrid systems based on graphene are of great interest. In such systems the interaction of Dirac fermions with other quasiparticles significantly changes properties of the system. In several papers it was studied the interaction between graphene and various systems: superconductors, thin films (2D) [5], nanowires (1D) [6], nanoholes and nanoperforations (0D) [7-9] and others. The authors of paper [10] considered theoretically a hybrid structure of superconducting islands array placed at the surface of graphene. A strong collective proximity effect in this structure was predicted due to a high mobility of massless carriers in graphene - Dirac fermions. Thus, superconductivity in graphene spreads at macroscopic lengths.

There were several experimental studies of the Josephson current through graphene in standard wide planar SNS junctions [11, 12]. However, until now experimental studies investigating this effect on the regular array structures are practically absent. There were only a few papers on studies this effect in non-regular structures [13]. Recently the first paper on studies of regular structure on graphene has been appeared [14]. Proximity effect in a system consisting of gold thin-films covered with an array of niobium nano-discs was studied in [15].

We report here on fabrication and investigation of the regular structure of superconducting islands on nanothin graphite. It is known that the surface of nanothin graphite also demonstrates the properties of Dirac fermions [8, 16-18]. The STM [19], cyclotron resonance [20] and Raman spectroscopy [21, 22] experiments showed that the surface layer of graphite is often represented as a graphene layer of an exceptional quality. In such system the Dirac fermions scattering is considerably lower than even in suspended graphene sheets. This graphene-on-graphite system is the most attractive for thin enough graphite samples to avoid the shunting of the surface graphene layers by the bulk

## 2. Materials and Methods

We obtained films of nanothin graphite using following technique: high-quality crystals of natural graphite were cleaned and then attached to the substrate using epoxy glue and then thinned down to tens of nanometers by adhesive tape [23, 24]. For making the surface of graphene clean we simply removed higher layers of graphite using adhesive tape. The flatness of the surface was controlled by optical microscope Carl Zeiss Axio Imager (with maximum magnification x1000) in differential interference contrast (DIC) regime. Usually we used 2-component 5-min epoxy glue UHU Schnellfest, which has excellent adhesive properties and flat surface without defects after drying. We usually used polycor ($Al_2O_3$) or silicon with 300-nm $SiO_2$ layer as a substrate. After thinning we were looking for a suitable graphite films (atomic thicknesses, dimensions more than 0.3x0.3 mm, with flat surface) using optical microscope in transmitting and reflecting modes. We controlled thicknesses of graphite films by light transmittance. It is known [25] that one graphene layer absorbs 2.3% of light independent of wavelength. We used translucent graphite films corresponding to thicknesses lower than 50 nm. The excess of films was cleaned off by razor blade.

After the suitable atomically thin graphite film was prepared we fabricated ohmic contacts to the sample. Graphite film was partially masked by whisker ribbons. Then Au contacts were manufactured on the laboratory laser ablation setup. The setup used the third harmonic (355 nm) of Nd:YAG laser with 0.2 J pulse energy, 10 ns pulse duration, and 6 Hz pulse repetition. Then the whisker mask was removed.

We tried to produce superconducting nanoislands on thin graphite by local tungsten (W) deposition using focused electron and ion beams of CrossBeam EsB 1540 (Carl Zeiss). We obtained the minimal diameter of nanoislands about 10 nm for electron beam deposition and 60-80 nm for ion beam deposition. However, we chose the ion beam deposition for our experiment because it was known from the literature [26-30] that in this case an alloy W-Ga-C is actually deposited. The alloy W-Ga-C temperature of superconducting transition $T_c$ is about 4-5 K that is much higher than $T_c$ of pure W. We produced the arrays from 10x7 to 40x20 nanoislands with the spacing from 100 to 300 nm (Fig. 1). For $T_c$ measurement W-Ga-C band was deposited. The measured resistance vs temperature dependence for W-Ga-C band is plotted on Insert of Fig. 2. We obtained $T_c \approx 5K$. To provide comparable measurements we fabricated two similar bridges, one containing W-Ga-C nanoislands (Fig.2) and another - without them (reference sample).

The measurements in magnetic field (up to 23.5 T) were performed with 10 MW Bitter magnet in the helium cryostat with possibility of cooling down to 1.7 K in Laboratoire National des Champs Magnétiques Intenses, Grenoble, France. The magnetic field was changed with rate 1-2 T/min. The samples resistance was measured by DC Current Reversal Method (with current value from 10 nA till 10 uA) for thermoelectric EMFs excluding. Keithley SourceMeter 2636A (with resolution of 20 fA) was used as a current source. Two nanovoltmeters Agilent 34420A (with resolution of 0.1 nV) were used for the synchronous voltage measurements on the sample containing W-Ga-C nanoislands and on the reference sample.

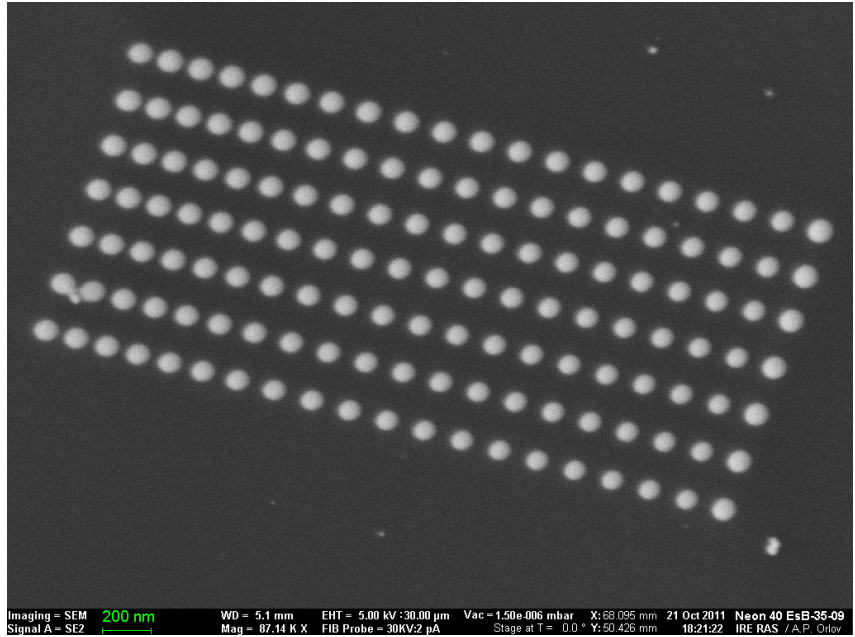

**Figure 1.** W-Ga-C nanoislands array on thin graphite. Nanoisland diameter is about 90 nm. Array spacing is 160 nm in horizontal row and 210 nm in vertical row.

## 3. Results and Discussion

Temperature dependences of the electro-resistance R (T) for both samples have similar character down to 5 K, whereas below 5 K the resistance of sample with nanoislands goes up considerably flatter than the resistance of reference bridge (Fig.3) indicating a proximity effect between the islands. Note, that the resistance of the sample with nanoislands is much higher than the resistance of the reference sample due to the sample damage by ion beam [31] during the deposition process. So, we compare the shapes of the corresponding R (T) curves and not the absolute values of the sample resistance.

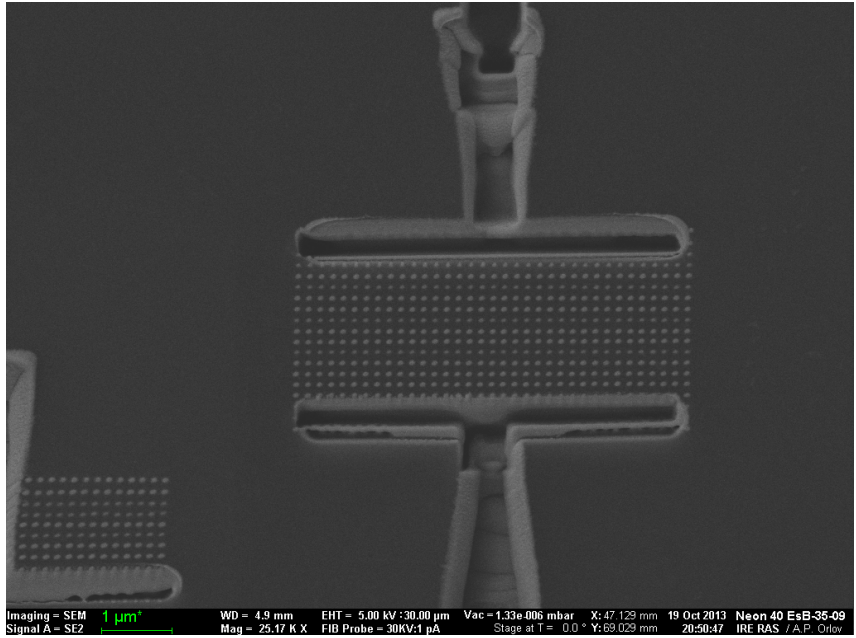

**Figure 2.** Graphite bridge with W-Ga-C nanoislands.

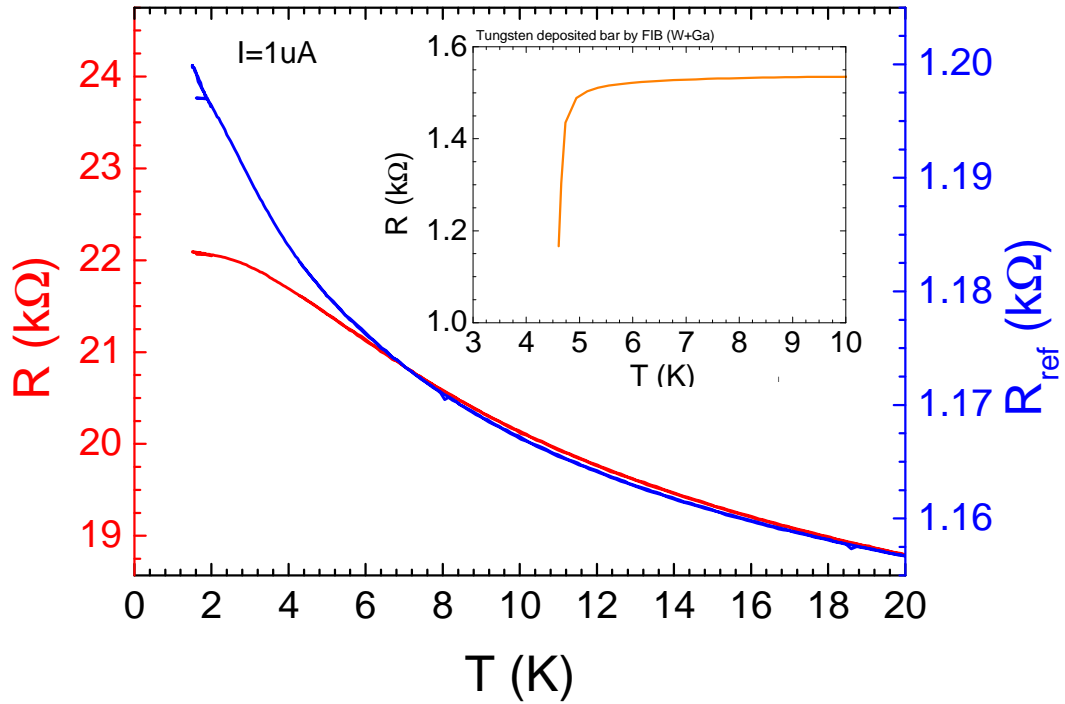

**Figure 3.** The resistance vs temperature dependences for samples containing the array of W-Ga-C nanoislands (red curve) and for reference sample without nanoislands (blue curve). The island diameter is 90 nm and the grating period is 170 nm. The lateral sizes of both bridges are 3x1.3 um. Inset: superconducting junction at the temperature dependence for W-Ga-C band.

Accordingly, the dependence of R (H) of the sample with islands first increases rapidly, which we attribute to the suppression of weak coupling between the islands, then proceeds to a negative magnetoresistance region, and finally goes to the usual dependence of R (H) similar to a reference sample (Fig.4). A region of negative magnetoresistance is attributed to the destruction of superconductivity in nanoislands at fields $H_{c1} < H < H_{c2}$ with a corresponding redistribution of the field and the weakening of its value between the islands, which leads to a decrease in the total sample magnetoresistance. Additionally, the suppression of superconductivity of nanoislands is also accomplished by an increase of normal carries, which become non-localized inside the island and contribute to total conductivity. A region of negative magnetoresistance has also been observed on the structures with irregular superconducting islands on graphene [13]. The increase of current through the sample leads to vanishing of a negative magnetoresistance region (Fig.5). At currents higher than of 500 nA the magnetoresistance of the graphite bridge with superconducting nanoislands looks similar to the magnetoresistance of the reference bridge. Also, the difference in behavior between the graphite sample with superconducting nanoislands and the reference sample vanished if the graphite thickness was increased up to one hundred of graphene monolayers.

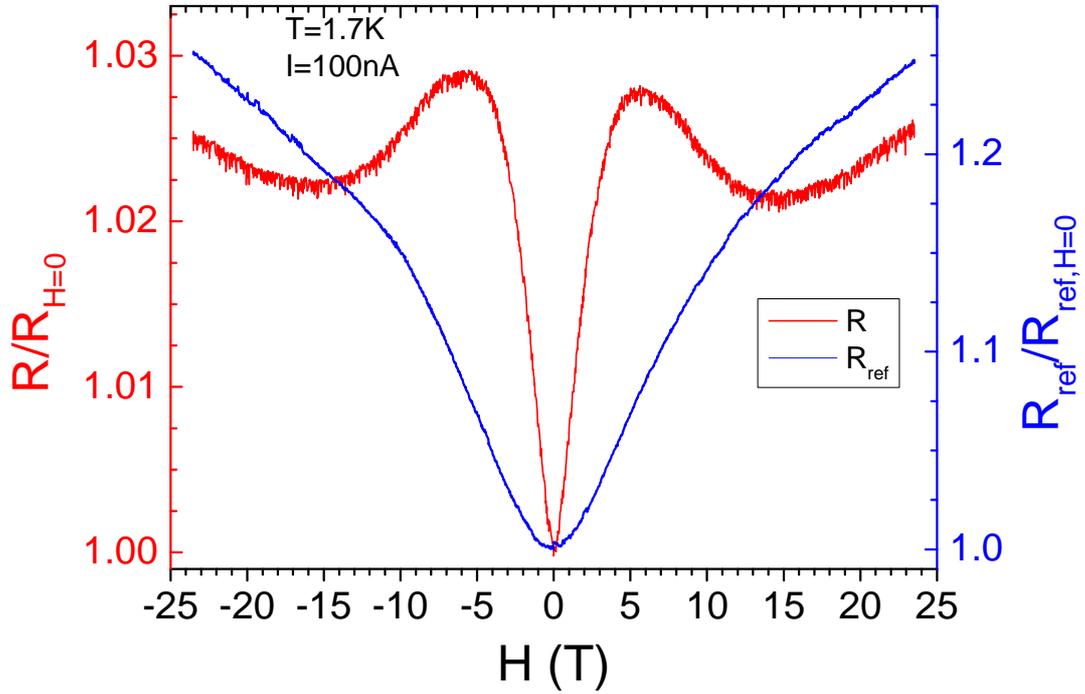

**Figure 4.** The magnetoresistance of the graphite bridge with superconducting nanoislands (red curve) and the reference bridge (blue curve).

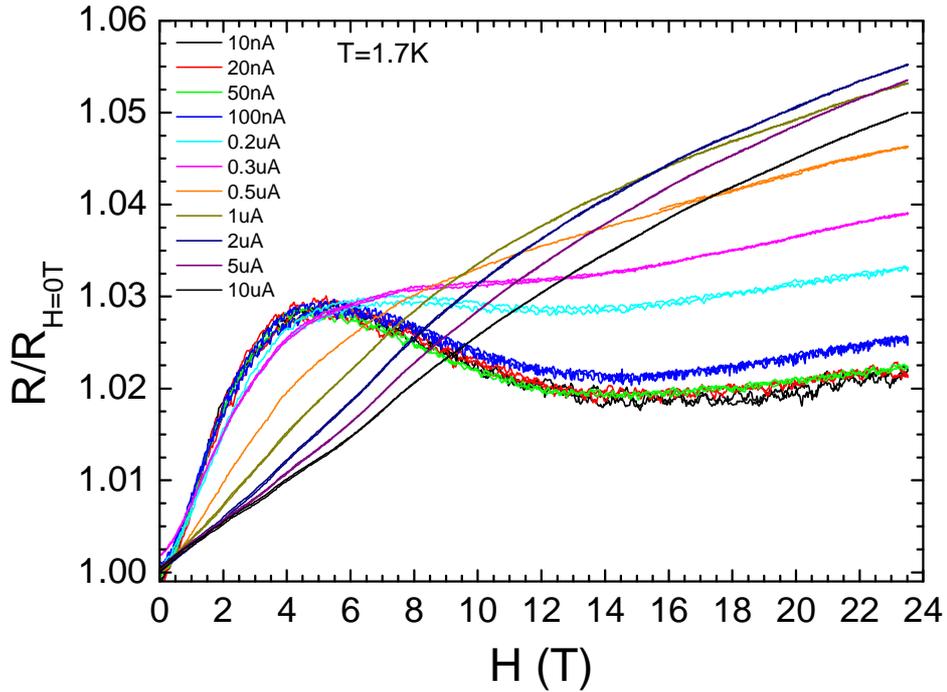

**Figure 5.** The magnetoresistance of the graphite bridge with superconducting nanoislands for different probe currents.

## 4. Conclusions

Graphene with superconductive nanoislands was studied. The periodic array of nanoislands of alloy W-Ga-C was fabricated on nanothin graphite by focused ion beam deposition. The nanoislands diameter was about 80 nm and the array spacing was from 100 to 300 nm. The resistance vs temperature dependence down to 1.7K and the magnetoresistance in field up to 24 T were measured both for the bridge containing nanoislands and for the reference bridge without islands. The resistance vs temperature curve below 5 K goes up considerably flatter for the bridge containing nanoislands. The resistance vs magnetic field curve shape for the bridge containing nanoislands differs considerably from the one for the bridge without nanoislands and has a region of negative magnetoresistance. Both differences were explained as the display of the proximity effect on the sample with superconductive nanoislands.

## Acknowledgements

The first author, Prof. Yuri Latyshev passed away in Moscow on 10th June 2014.

The authors thank A.A. Sinchenko for helpful discussions and V.A. Shakhunov for sample contacts manufacturing. The work has been supported by RFBR (grants No. 14-02-01126-a, 14-02-01166-a and 13-02-00927-a), by the programs of RAS, and by the Laboratoire National des Champs Magnétiques Intenses (Grenoble).

The work was presented on the 4th International scientific conference state-of-the-art trends of scientific research of artificial and natural nanoobjects (STRANN '14) 22-25, April 2014 Saint-Petersburg, Russia [32].